# Giant magnetostriction and nonsaturating electric polarization up to 60 T in the polar magnet CaBaCo4O7

Yi-Sheng Chai[1*], Jun-Zhuang Cong[2], Jin-Cheng He[2], Dan Su[2], Xia-Xin Ding[3], John Singleton[3], Vivien Zapf[3], and Young Sun[1,2†]

[1]Center of Quantum Materials and Devices and Department of Applied Physics, Chongqing University, Chongqing 401331, China

[2]Beijing National Laboratory for Condensed Matter Physics and Beijing Advanced Innovation Center for Materials Genome Engineering, Institute of Physics, Chinese Academy of Sciences, Beijing 100190, China

[3]National High Magnetic Field Laboratory, Los Alamos National Laboratory, Los Alamos, New Mexico 87545, USA

[*]yschai@cqu.edu.cn

[†]youngsun@cqu.edu.cn

Abstract

Giant magnetostriction in insulating magnetic materials is highly required for applications but is rarely observed. Here we show that giant magnetostriction (> 1500 ppm) can be achieved in an insulating transition metal oxide CaBaCo$_4$O$_7$ where the ferrimagnetic ordering at $T_C$ ~ 62 K is associated with a huge change in the lattice. Moreover, because this material is pyroelectric with a non-switchable electric polarization ($P$), the giant magnetostriction results in a record-breaking magnetoelectric effect – a gigantic change of electric polarization ($\Delta P$ ~ 1.6 μC/cm$^2$) in response to the applied magnetic field up to 60

T. Geometric frustration as well as the orbital instability of $Co^{2+}/Co^{3+}$ ions is believed to play a crucial role in the giant magnetostriction. Our study provides new insights on how to achieve both giant magnetostriction and pronounced magnetoelectric effect in insulating transition metal oxides.

**I. INTRODUCTION**

Magnetostriction refers to the deformation of a material in response to a change in its magnetization [1]. Materials with giant magnetostriction have been widely used in a broad range of devices such as magnetostrictive actuator, transducer, electric generator, motor, sensor, etc [1]. All magnetic materials exhibit magnetostriction to some degree. However, giant magnetostriction is generally observed in some elementary rare-earth metals and a few alloys combining rare-earth elements and transition metals (Fe, Ni, and Co) [1,2]. In order to develop alternative magnetostrictive materials with reduced eddy current and lower cost, oxide-based insulating magnetostrictive materials have also been studied for many years. Although notable magnetostriction was reported in some magnetic perovskite and spinel oxides [3-7], the amplitude of magnetostriction is not as high as that of the rare-earth alloys. Therefore, insulating magnetostrictive materials with giant magnetostriction remain a big challenge.

On another side, the magnetoelectric (ME) effects, *i.e.*, the mutual control of electric polarization ($P$) by a magnetic field ($H$) and magnetization ($M$) by an electric field, have attracted enormous attention due to their intriguing physics as well as potential applications [8-10]. A large number of ME materials have been discovered in the past decade [11-14]. Nevertheless, the amplitude of magnetic-field tunable electric polarization in single-phase

ME materials is quite low [15-18], usually less than 0.1 μC/cm$^2$. A relatively large value of Δ$P$ (~ 0.3 – 0.5 μC/cm$^2$) was ever reported in multiferroic CaMn$_7$O$_{12}$ (ref. 19) and GdMn$_2$O$_5$ (ref. 20) with the established mechanisms of spin-driven ferroelectricity. In order to further enhance the tunable electric polarization in ME materials, new principles and materials are required. To date, the most pronounced ME effects are observed in ferroelectric/ferromagnetic composites [21] where the ferromagnetic component with a large magnetostrictive coefficient is a key ingredient to produce the large ME effects via interfacial elastic coupling. Similarly, if giant magnetostriction can be introduced into a single-phase polar magnet (either ferroelectric or pyroelectric), a significant ME effect in bulk form could be expected. Here, we demonstrate that this strategy can be indeed accomplished in CaBaCo$_4$O$_7$ which is a unique polar magnet showing giant magnetostriction.

CaBaCo$_4$O$_7$ was first synthesized in 2009 by Caignaert *et al* [22]. Its crystalline structure consists of a 1:1 stacking of pseudo-kagome and pseudo-triangular layers of CoO$_4$ tetrahedra along the *c* axis, as shown in Fig. 1(a). CaBaCo$_4$O$_7$ undergoes a structural phase transition from hexagonal (space group *P6$_3$mc*) to orthorhombic (space group *Pbn2$_1$*) at 450 K [23]. The polar space group *Pbn2$_1$* indicates that it could be ferroelectric in the low temperature phase. Previous studies on polycrystalline CaBaCo$_4$O$_7$ claimed that it is multiferroic with ferrimagnetic ordering and spin-assisted ferroelectricity [24,25]. Later, measurements on single-crystal samples suggest that it is pyroelectric rather than ferroelectric because the electric polarization is non-switchable [26,27]. A significant magnetic-field tunable electric polarization (Δ$P$ ~ 0.8 μC/cm$^2$) was observed near the magnetic ordering temperature $T_C$, but the underlying mechanism was unclarified yet.

Meanwhile, $CaBaFe_4O_7$, a derivative material of $CaBaCo_4O_7$, also attracted much attention because it shows strong ME effects at relatively higher temperatures [28,29].

In this work, we have investigated thermal expansion, magnetostriction, and the ME effect on single-crystal samples of $CaBaCo_4O_7$. The amplitude of the tunable electric polarization does not saturate up to 60 T, yielding an unprecedentedly high value of $\Delta P \sim$ 1.6 $\mu C/cm^2$. These results not only demonstrate remarkable magnetostriction in an insulating oxide but also clarify that the gigantic tunability of electric polarization is due to the giant magnetostriction in a polar magnet.

## II. EXPERIMENTS

Single crystals of $CaBaCo_4O_7$ were grown using the floating zone technique in a mirror furnace similar to that described in previous work [8]. The largest sample has a size of $1.0 \times 1.0 \times 0.6$ mm$^3$. The crystalline quality was checked by x-ray Laue diffraction and found to be single phase. The temperature dependence of magnetization in low magnetic field was measured in a magnetic property measurement system (MPMS-XL, Quantum Design). The pyroelectric current was measured by a Keithley 6517B electrometer using a home-made probe in a Cryogen-free Superconducting Magnet System (Oxford Instruments, TeslatronPT). For the pyroelectric current measurements, the sample was poled in an electric field of ±10 kV/cm from 200 to 5 K. After removing the poling electric field and releasing space charges for at least 30 min, the pyroelectric current was recorded with warming at a constant rate of 1 K/min. The magnetization and electric polarization up to 60 T were measured in the pulsed high magnetic field facility at Los Alamos National Laboratory. A capacitance dilatometer [30] was employed to measure the linear thermal expansion and magnetostriction in a Cryogen-free Superconducting Magnet System (Oxford Instruments, TeslatronPT).

## III. RESULTS AND DISCUSSION

Figure 1(b) shows the temperature dependence of magnetization along the *b*-axis (magnetic easy-axis) measured in a low magnetic field (500 Oe). A paramagnetic to ferrimagnetic transition occurs at $T_C \sim 62$ K, consistent with previous reports [25]. Accompanying with this magnetic phase transition, a sharp pyroelectric current peak along the *c*-axis appears (Fig. 1(c)), implying a big change in electric polarization. We note that the pyroelectric peak does not reverse its sign with positive or negative poling electric field, which confirms that $CaBaCo_4O_7$ is pyroelectric rather than ferroelectric. As seen in Fig. 2(c), the amplitude of $\Delta P$ (> 0.7 μC/cm$^2$) is apparently larger than that of many magnetically driven type-II multiferroics where the inverse Dzyaloshinskii–Moriya (DM) interaction [26] or exchange striction mechanism [27] plays a dominating role. Therefore, an alternative mechanism could be responsible for the magnetic order induced electric polarization in $CaBaCo_4O_7$.

The thermal expansion measurements clearly demonstrate a significant change (> 1500 ppm) in the lattice along the *c*-axis across the ferrimagnetic phase transition, shown in Fig. 2(a), indicating a strong magneto-structural correlation in $CaBaCo_4O_7$. In terms of the lattice change of the *c*-axis associated with the ferrimagnetic ordering, the induced electric polarization ($\Delta P$) is a natural consequence of the structural change along the polar direction. One peculiar feature of $CaBaCo_4O_7$ is that the structural transition near $T_C$ is sensitive to external magnetic fields. As shown in Fig. 2(a), the thermal expansion along the *c*-axis strongly depends on the applied magnetic field in the *ab*-plane. In zero magnetic field, a single sharp structural transition occurs at $T_C$=62 K. When an in-plane magnetic field is applied, the structural phase transition shifts rapidly to higher temperatures, and is broadened and split into two steps in high magnetic fields. Correspondingly, the sharp pyroelectric peak in zero field shifts to a higher temperature with increasing magnetic field and splits into two peaks (marked with $T_C$ and $T^*$, respectively) in high magnetic fields, shown in Fig. 2(b). The transition at $T_C$ is related to magnetic ordering, but it is not clear what is the origin of the minor transition at $T^*$. As seen in Fig. 2d, both $T_C$ and $T^*$ shift almost linearly with magnetic field. The peak at $T^*$ is broader and shift faster than the peak at $T_C$.

After integrating the pyroelectric current with time, we obtained the electric polarization along the *c*-axis as a function of temperature, shown in Fig. 2(c). Due to the shift of pyroelectric peaks, the temperature dependence of electric polarization changes significantly with applied in-plane magnetic field, resulting in a pronounced ME effect above $T_C$. In other words, the ME effect in $CaBaCo_4O_7$ mainly happens in the paramagnetic state rather than the magnetic ordering state. This feature is in strong contrast to spin-driven type-II multiferroics where the ME effect normally occurs in the magnetic ordering state below $T_C$. Thus, there is a distinct mechanism underlying the ME effect in $CaBaCo_4O_7$.

As the structural phase transition temperature is sensitive to external magnetic fields, large magnetostriction would be expected in the vicinity of the phase transition. Fig. 3(a) shows the magnetostriction ($\Delta L/L$) behavior along *c* axis of $CaBaCo_4O_7$ at selected temperatures around $T_C$. At 60 K which is slightly below $T_C$, the magnetostriction is relatively small, ~ 160 ppm for 7 T magnetic field. At $T_C$=62 K, the magnetostriction is remarkable, reaching ~ 1500 ppm for 7 T. This value is comparable to the maximum magnetostriction (1500 – 2000 ppm) of the famous TbDyFe alloys [1]. With further increasing temperature, the magnetostriction decays, but remains a high value up to 70 K. These results demonstrate that giant magnetostriction can be indeed obtained in insulating transition metal oxides.

Since $CaBaCo_4O_7$ is pyroelectric, the lattice change caused by magnetostriction will certainly induce a change in electric polarization. We have measured the direct ME effect (magnetic field control of electric polarization) at selected temperatures. As shown in Fig. 3(b), the change of electric polarization ($\Delta P = P(\mu_0 H) - P(0\ T)$) along $c$-axis as a function of applied $ab$-plane magnetic field is most significant at $T_C$ and decays with increasing or decreasing temperature, very similar to the behavior of magnetostriction (Fig. 3(a)). Apparently, there is a close correlation between the magnetostriction and ME effect as expected for a polar magnet.

Because $CaBaCo_4O_7$ is ferrimagnetic below $T_C$, its magnetization does not saturate up to 7 T. Correspondingly, the magnetostriction as well as the induced electric polarization do not show a saturation with increasing magnetic field. It would be very interesting to seek the ultimate limit of magnetic-field tunable electric polarization. To this end, we have performed the magnetization and magnetoelectric current measurements up to 60 T using the pulsed high magnetic field facility at Los Alamos National Laboratory. Figure 4(a) shows the isothermal magnetization curves at selected temperatures. To our surprise, the magnetization increases smoothly and does not saturate up to 60 T at all the temperatures studied. A spin-flop transition expected for ferrimagnets does not show up till this field. The non-saturating magnetization indicates that the giant magnetostriction in $CaBaCo_4O_7$ could be non-saturating even at 60 T. Unfortunately, we are not able to directly measure the magnetostriction under the pulsed high magnetic fields in this study.

Figure 4(b) presents the induced electric polarization ($\Delta P$) as a function of high magnetic field up to 60 T at selected temperatures. The electric polarization is obtained by integrating the magnetoelectric current with time. At $T_C$, the largest tunable electric polarization ($\Delta P$) reaches ~ 1.6 $\mu C/cm^2$, which sets a new record in ME materials. Similar to the magnetization, the induced electric polarization does not saturate at 60 T and a larger $\Delta P$ would be expected at higher magnetic fields.

The above results demonstrate remarkable properties in the polar magnet $CaBaCo_4O_7$: giant magnetostriction and record-breaking ME effect. Meanwhile, our study clarifies several controversial issues on $CaBaCo_4O_7$. First, it is confirmed that $CaBaCo_4O_7$ is pyroelectric rather than ferroelectric as several reports claimed. Second, the significant ME effect in $CaBaCo_4O_7$ mainly happens near the magnetic phase transition in the paramagnetic state. The ME effect in the magnetic ordering state is actually weak, which is in contrast to many spin-induced multiferroics. Third, the mechanism of the pronounced ME effect is related to the giant magnetostriction along the polar axis in a polar magnet. A recent study [31] using *ab initio* calculation suggested that the exchange striction upon ferrimagnetic ordering is strong enough to produce a giant change in electric polarization in $CaBaCo_4O_7$. This mechanism is distinct to those well-established models in single-phase ME materials. The key point is to introduce giant magnetostriction in a polar magnet so that a significant ME effect would be naturally obtained.

The origin of giant magnetostriction in $CaBaCo_4O_7$ is believed to be closely related to the states of Co ions. In previous studies, large magnetostriction was found in several cobalt oxides including $CoFe_2O_4$ [3] and $La_{1-x}Sr_xCoO_3$ [4]. In the former, the magnetostriction is associated with the high spin state of $Co^{2+}$ ions[3]. In the latter, the magnetostriction is ascribed to the spin state transition of $Co^{3+}$ ions under the applied magnetic fields [7]. Interestingly, $CaBaCo_4O_7$ exhibits charge ordering with the stoichiometric formula $CaBaCo_2^{2+}Co_2^{3+}O_7$ so that both $Co^{2+}$ and $Co^{3+}$ ions exist. It is likely that both the unquenched orbital angular momentum of $Co^{2+}$ and the orbital instability of $Co^{3+}$ contribute to the giant magnetostriction. Meanwhile, the structure of $CaBaCo_4O_7$ comprises interleaved kagome and triangular layers of $CoO_4$ tetrahedra. The geometric frustration intrinsic to kagome and triangular lattices is lifted by a strong buckling of $CoO_4$ tetrahedra, which induces the appearance of a ferrimagnetic state. The highly distorted geometry of the $CoO_4$ tetrahedra is sensitive to external stimuli and can be significantly modified by varying temperature and magnetic field. All these factors may jointly result in a giant magnetostriction in $CaBaCo_4O_7$.

## IV. CONCLUSION

In summary, a non-saturating giant ME effect ($\Delta P \sim 1.6$ μC/cm$^2$) up to 60 T is observed in the pyroelectric ferrimagnetic $CaBaCo_4O_7$. The underlying physics is related to the giant magnetostriction at temperatures slightly above the magnetic phase transition. The sensitivity of the lattice to external magnetic fields is ascribed to the orbital instability of $Co^{2+}/Co^{3+}$ ions as well as the geometric frustration. In the future, people may look for giant magnetostriction in cobalt oxides with geometric frustration. Furthermore, gigantic ME effect can be achieved in single-phase materials by a strategy of introducing pronounced magnetostriction into a polar magnet.

**ACKNOWLEDGEMENTS**

This work was supported by Beijing Natural Science Foundation (Grant No. Z180009) and the National Natural Science Foundation of China (Grant Nos. 51725104, 11674384, 11974065). A portion of this work was performed at the National High Magnetic Field Laboratory, which was supported by National Science Foundation Cooperative Agreement No. DMR-1157490 and the State of Florida.

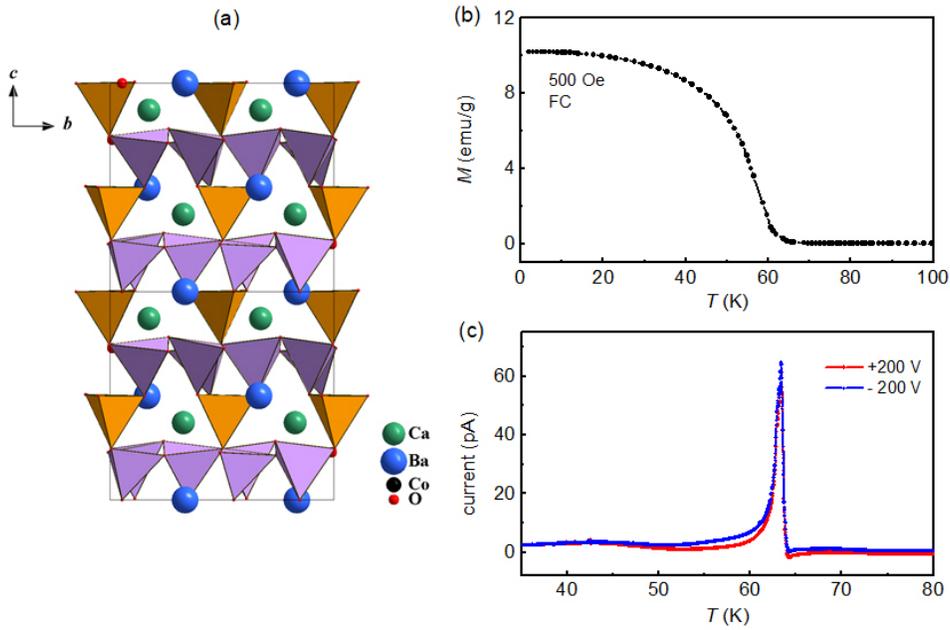

**Fig. 1** (a) The crystalline structure of CaBaCo$_4$O$_7$. It consists of a 1:1 stacking of kagome and triangular layers of CoO$_4$ tetrahedra along the *c* axis. (b) Temperature dependence of magnetization along the *b*-axis. A ferrimagnetic transition occurs at $T_C \sim 62$ K. (c) Pyroelectric current along the *c*-axis as a function of temperature. A sharp pyroelectric peak appears at the magnetic ordering temperature $T_C$. The pyroelectric current does not reverse its sign after a negative poling, suggesting that CaBaCo$_4$O$_7$ is pyroelectric rather than ferroelectric.

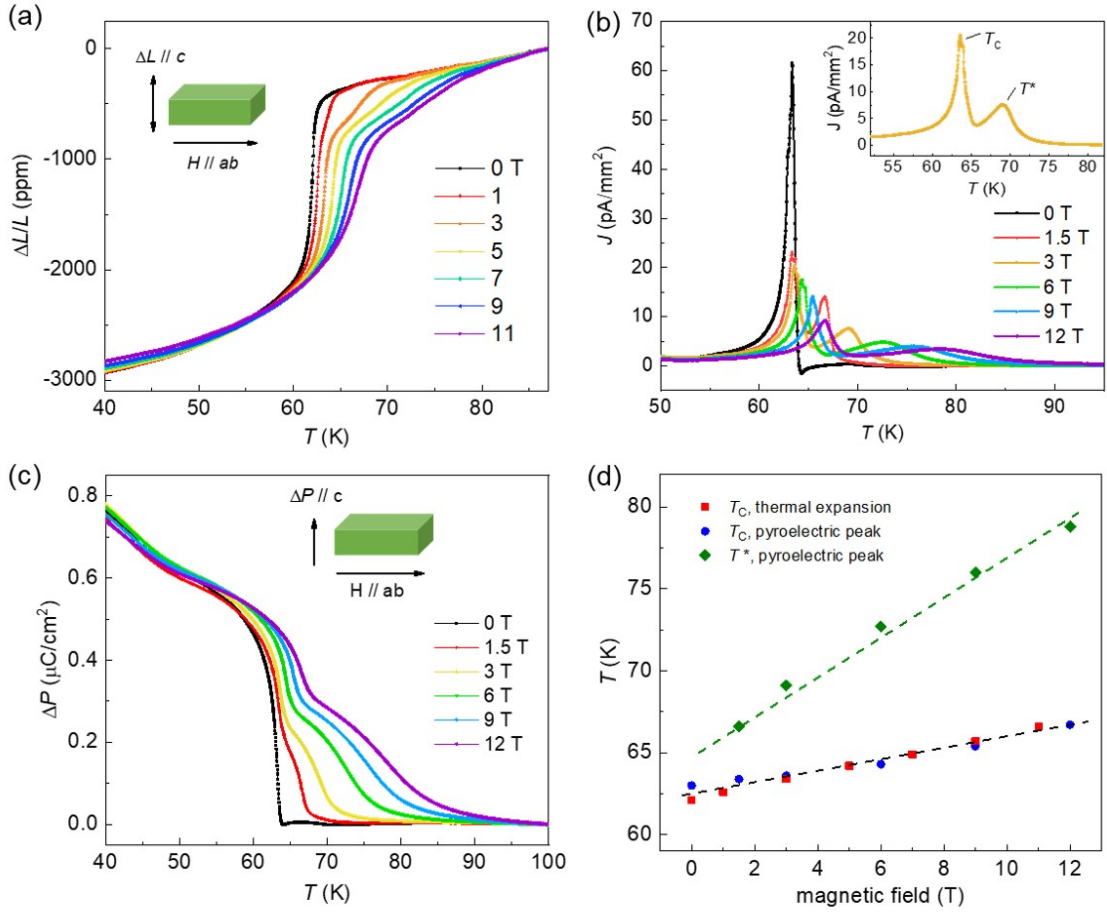

**Fig. 2** (a) Temperature dependence of thermal expansion along the *c*-axis of $CaBaCo_4O_7$ measured in a series of magnetic fields applied in *ab* plane. (b) Temperature dependence of pyroelectric current measured in several magnetic fields. The inset shows that there are two pyroelectric peaks at $T_C$ and $T^*$, respectively. (c) Temperature dependence of electric polarization along the *c*-axis under several magnetic fields. (d) The critical temperatures $T_C$ and $T^*$ as a function of magnetic field.

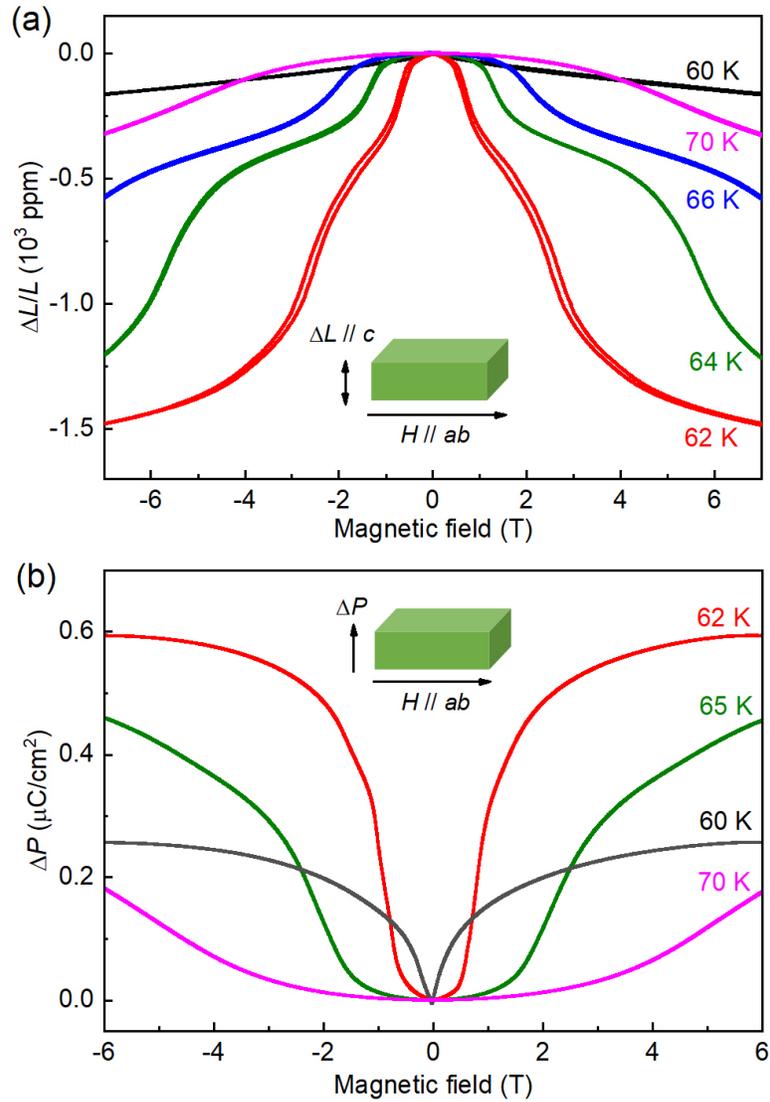

**Fig. 3** (a) Magnetostriction along the *c*-axis of CaBaCo$_4$O$_7$ with *ab* plane magnetic field at selected temperatures. A giant magnetostriction is observed at temperatures close to $T_C$. (b) Magnetic field tuning of electric polarization at selected temperatures. The electric polarization is measured along the *c*-axis.

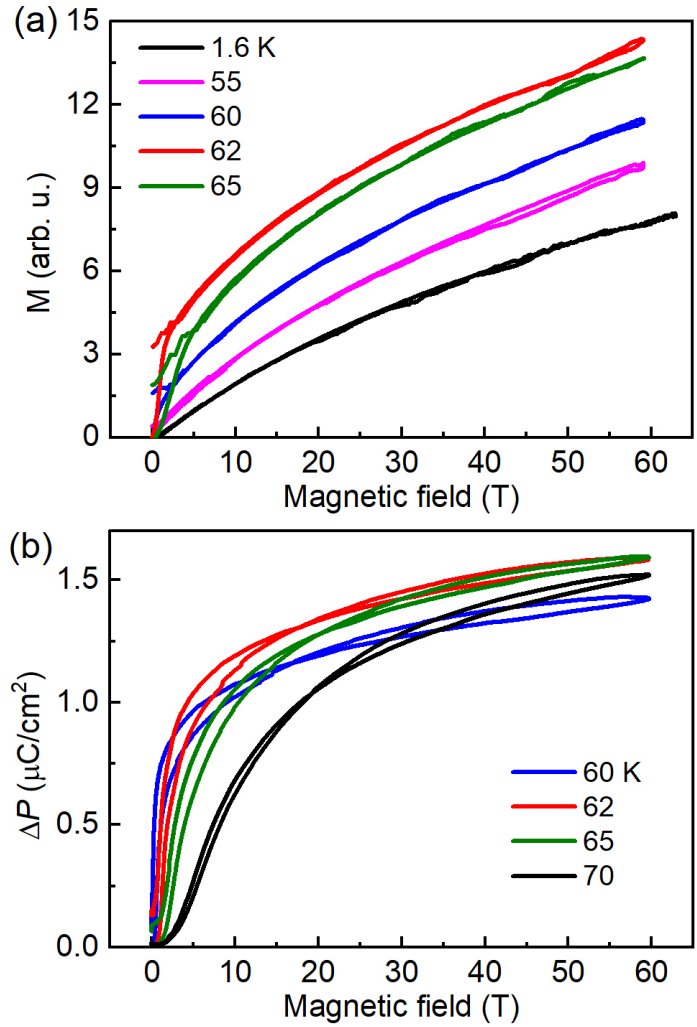

**Fig. 4** (a) The *M-H* curves along *b*-axis of CaBaCo$_4$O$_7$ at selected temperatures. The magnetization does not saturate up to 60 T. (b) Induced electric polarization as a function of magnetic field at selected temperatures. The electric polarization does not saturate up to 60 T.